\documentclass[12pt,aps,prb,preprint]{revtex4}
\usepackage{amsmath,amssymb}
\usepackage{graphicx}   
\usepackage{bm}

\begin{document}

\title{Spherical volume averages of static electric and magnetic fields using Coulomb and Biot-Savart laws}

\author{Ben Yu-Kuang Hu}
\email{byhu@uakron.edu} \affiliation{Department of Physics,
University of Akron, Akron, OH~44325-4001}

\date{\today}

\begin{abstract}
We present derivations of the expressions for the spherical volume
averages of static electric and magnetic fields that are virtually
identical. These derivations utilize the Coulomb and Biot-Savart
laws, and make no use of vector calculus identities or potentials.
\end{abstract}
\maketitle

\section{Introduction}
The average of static electric or magnetic fields over a sphere
has been used to obtain important results, such as the macroscopic
electric field inside a dielectric\cite{griffiths} and the
presence of $\delta$-function electric (magnetic) fields at the
position of electric (magnetic) point dipoles.\cite{jackson}  In
both electric and magnetic cases, if the sources of the fields are
outside the averaging sphere, the average field equals the value
of the field at the center of the sphere, whereas for sources
inside the averaging sphere, the average electric and magnetic
fields are proportional to the electric and magnetic dipole
moments of the sources, respectively.

Textbooks typically only treat the electric field
case.\cite{electric_only} Only a handful of authors
\cite{jackson,both} treat both electric and magnetic cases,
probably because the magnetic field case is considered to be
``tough" by undergraduate standards,\cite{griffiths1} since the
derivations typically employ the magnetic vector potential and
vector calculus identities.   This paper presents derivations of
both the electric and magnetic field cases that are virtually
identical and are elementary, in the sense that they rely on
Coulomb's and the Biot-Savart laws and make no use of vector
calculus identities or potentials.

Coulomb's law states that the electric field for a unit point
charge is
\begin{equation}
{\bm G}(
\bm r) = k\frac{\bm r}{r^3},
\end{equation}
where $\bm r$ is a displacement vector from the point charge, $r =
\vert\bm r\,\vert$ and  $k = (4\pi\epsilon_0)^{-1}$ and $1$ in SI
and Gaussian units, respectively.  In terms of $\bm G$, the
electric field the electric field ${\bm E}$ at point $\bm s$ due
to a static charge distribution $\rho(\bm r')$ is
\begin{align}
\bm E(\bm s) &= \int d\bm r'\ \rho(\bm r')\, \bm G(\bm s-\bm r') =
-\int d\bm r'\ \rho(\bm r')\,\bm G(\bm r'-\bm s).\label{eq:2}
\end{align}

Define $\overline{\bm G_r}(\bm r')$ to be the {\em average of $\bm
G$ over a spherical surface of radius $r$ around a point $\bm
r'$}, as shown in Fig.~1. An apparently little-known result that
is critical in following discussion is (see the appendix for
derivations)
\begin{equation}
\overline{\bm G_r} (\bm r')= k\frac{\bm r'}{r'^3}\;\Theta(r' - r)
\label{eq:4}
\end{equation}
where $\Theta$ is the Heavyside step function.\cite{heavyside}
Eq.~(\ref{eq:4}) implies that the average of the electric field
over the surface of the sphere for a point charge outside the
sphere ($r' > r$) equals the electric field at the center of the
sphere, whereas a point charge inside a sphere ($r' < r$)
contributes {\em zero} to the average of the electric field over
the surface of the sphere.  In a sense, this result is the
antithesis of the integral form of Gauss' law,\cite{difference}
but it is not as general as Gauss' law because it applies only to
spherical surfaces.

\section{Volume averages of static electric fields}

Let $\langle\bm E\rangle_R$ to be average of the electric field
$\bm E$ about a spherical {\em volume} of radius $R$ and
$\overline{\bm E_r}(\bm 0)$ be the average electric field over a
spherical shell of radius $r$, both centered around the origin.
Then,
\begin{eqnarray}
\langle\bm E\rangle_R &\equiv& \frac{3}{4\pi R^3}\int_{r=0}^R\!
d{\bm r}\ \bm E(\bm r) = \frac3{R^3}\int_{0}^R dr\ r^2\;
\overline{\bm E_r}(\bm 0). \label{eq:5}
\end{eqnarray}
For a given charge distribution $\rho(\bm r')$, $\overline{\bm
E_r}(\bm 0)$ can be obtained by setting $\bm s = \bm 0$ and
averaging both sides of Eq.~(\ref{eq:2}) over a spherical shell of
radius $r$, yielding
\begin{align}
\overline{\bm E_r}(\bm 0) &= -\int d\bm r'\ \rho(\bm r')\,
\overline{\bm G_r}(\bm r').\label{eq:6}
\end{align}

\subsection{Single point charge}

We first examine the case of a single point charge. Because
electric fields obey the principle of linear superposition, what
holds for a single point charge is generalizable to cases of many
charges and/or continuous charge distributions.  Substituting the
charge distribution for a point charge $q$ at $\bm d$, $\rho(\bm
r') = q\,\delta(\bm r'-\bm d)$, into Eq.~(\ref{eq:6}) gives
$\overline{\bm E_r}(\bm 0) = -q\overline{\bm G_r}(\bm d)$.
Substituting this into Eq.~(\ref{eq:5}) and using Eq.~(\ref{eq:4})
gives
\begin{equation}
\langle\bm E\rangle_R = -\frac{3kq\bm d}{d^3R^3}\int_{0}^{R} dr\
r^2\;\Theta(d - r).\label{eq:6a}
\end{equation}
If the point charge is outside the averaging sphere of radius $R$,
then $d > R$ and the integral in Eq.~(\ref{eq:6a}) equals $R^3/3$.
Hence, $\langle\bm E\rangle_R^{\rm (out)} = -kq\bm d/d^3 = \bm
E(\bm 0)$, the electric field at the center of the sphere due to
the point charge.  On the other hand, if the point charge is
inside the averaging sphere, then $d < R$ and the integral in
Eq.~(\ref{eq:6a}) equals $d^3/3$, in which case $\langle\bm
E\rangle^{\rm(in)}_R = -k\bm p/R^3$, where $\bm p = q\bm d$ is the
dipole moment of the point charge relative to the center of the
sphere.

\subsection{Arbitrary charge distribution}

We now obtain these results more rigorously for an arbitrary
charge distribution $\rho(\bm r')$.  Substituting
Eqs.~(\ref{eq:4}) and (\ref{eq:6}) into Eq.~(\ref{eq:5}) yields,
\begin{align}
\langle{\bm E}\rangle_R &= -\frac{3}{R^3}\int_0^R dr\ r^2
\left[\int_{r'=r}^{\infty} d\bm r'\ \frac{k\,\rho(\bm r')\;\bm
r'}{r'^3}\right].
 \label{eq:7}
\end{align}
where the effect of the $\Theta$ function in Eq.~(\ref{eq:4}) is
to restrict the $\bm r'$ integration to the region $\vert \bm
r'\vert>r$.  We now consider separately the contribution of the
charges outside and inside the sphere.

\subsubsection{Sources outside the sphere}  The contribution
due to charges outside the sphere of radius $R$ corresponds to
$\int_{r'=R}^\infty d\bm r'$ in Eq.~(\ref{eq:7}), which makes the
term in square parentheses independent of $r$.  The integration
over $r$ gives $R^3/3$, yielding
\begin{equation}
\langle \bm E\rangle^{\rm (out)}_{R} = -\int_{r'=R}^\infty\! d\bm
r'\ \frac{k\,\rho(\bm r')\;\bm r'}{r'^3} = \bm E^{\rm (out)}(\bm
0)\label{eq:average_out}
\end{equation}
the electric field at the origin. Therefore, in general,
$\langle\bm E\rangle_R^{\rm(out)} = \bm E^{\rm(out)}$ at the
center of the sphere.

\subsubsection{Sources inside the sphere}

Sources inside the sphere of radius $R$ correspond to $\vert\bm
r'\vert < R$ in Eq.~(\ref{eq:7}).  This gives
\begin{align}
\langle\bm E\rangle_R^{\rm(in)} &= -\frac{3}{R^3}\int_{0}^R dr\ r^2 \int_{r'=r}^R d\bm r'\ \frac{k\,\rho(\bm r')\;\bm r'}{r'^3} \nonumber\\
&= -\frac{3k}{R^3}\int_{r'=0}^R d\bm r'\ \frac{\rho(\bm r')\; \bm r'}{r'^3}\int_{0}^{r'} dr\ r^2  \nonumber\\
&= -\frac{k}{R^3}\int_{r'=0}^R d\bm r'\; \rho(\bm r')\; {\bm r'} =
-\frac{k\,\bm p}{R^3}, \label{eq:average_in}
\end{align}
where $\bm p = \int d\bm r'\ \rho(\bm r')\,\bm r'$ is the dipole
moment relative to the center of the sphere, and we have used
$\int_0^R dr \int_r^R dr' = \int_0^R dr' \int_0^{r'} dr$, as shown
in Fig.~2.

\section{Magnetic field case}

A static magnetic field $\bm B$ is related to the charge current
density $\bm J$ by the Biot-Savart law,
\begin{equation}
\bm B(\bm s) = \int d\bm r'\ \bm J(\bm r')\bm\times \bm G(\bm
s-\bm r'),\label{eq:10}
\end{equation}
where $k = \mu_0/(4\pi)$ or $c^{-1}$ for SI or gaussian,
respectively. Comparison of Eq.~(\ref{eq:10}) with
Eq.~(\ref{eq:2}) shows that the derivation for the magnetic field
case can be copied wholesale from the electric field case by
substituting $\bm E \rightarrow \bm B$ and $\rho\rightarrow \bm
J\bm\times$ at every step. (The only point of caution is that the
cross product anti-commutes, so care must be taken to preserve the
order of $\bm J$ and $\overline{\bm G}$ or $\bm r'$.) Using these
substitutions in Eq.~(\ref{eq:average_out}) gives, for current
sources outside the averaging sphere,
\begin{align}
\langle\bm B\rangle_R^{\rm (out)} &= -\int_{r'=R}^\infty d\bm r'\
\frac{k\,\bm J(\bm r')\times\bm r'}{r'^3} = \bm B(\bm 0),
 \label{eq:9}
\end{align}
the magnetic field at the center of the sphere. For current
sources inside the averaging sphere, using the substitutions in
Eq.~(\ref{eq:average_in}) gives
\begin{align}
\langle\bm B\rangle_R^{\rm (in)} &= -\frac{k}{R^3}\int_{r'=0}^R
d\bm r' \bm J(\bm r')\times {\bm r'} = \frac{\,2k\bm m}{R^3},
\end{align}
where $\bm m = \frac{1}{2}\int d\bm r'\ \bm r' \bm \times \bm
J(\bm r')$ is the magnetic dipole moment.\cite{jackson1}
\appendix
\section{Derivation of Eq.~(\ref{eq:4})}

By definition,
\begin{equation}
\overline{\bm G_r}(\bm r') = \frac1{4\pi r^2}\int dA''\ \bm G(\bm
r''),
\end{equation}
where the integral is over the spherical shell of radius $r$
around $\bm r'$.  Choose $\bm r' = r'\hat{\bm z}$, where $\hat{\bm
z}$ is the unit vector in the $z$-direction.  By azimuthal
symmetry, $\overline{\bm G_r}(r'\hat{\bm z}) =
\overline{G_r}_{,z}\hat{\bm z} = (4\pi r^2)^{-1}\int dA''\,G_z(\bm
r'') \hat{\bm z}$, where $G_z$ is the $z$-component of $\bm G$. By
dividing the spherical surface into strips $dA'' = r^2
\sin\theta\, d\theta$ (see Fig.~3), $\overline{G_r}_{,z} =
\frac{1}{2} \int_0^{\pi} d\theta\ \sin\theta\ G_z(\bm r'')$.
Letting $\mu = \cos\theta$, and using $G_z(\bm r'') = k r_z''/|\bm
r''|^3$ gives
\begin{align}
\overline{G_r}_{,z} &= \frac{k}2 \int_{-1}^1 d\mu\ \frac{r\mu+r'}{(r^2 + r'^2 + 2 r' r \mu)^{3/2}}\nonumber\\
&= -\frac{k}{2}\frac\partial{\partial r'}\left[\int_{-1}^1 d\mu\ \frac1{(r^2 + r'^2 + 2 r' r \mu)^{1/2}}\right]\nonumber\\
&= -\frac{k}{2}\frac\partial{\partial r'}\left[\frac{\sqrt{r^2 +
r'^2 + 2 r' r \mu}}{r'r}\right]_{\mu=-1}^{1}
= -\frac{k}{2}\frac{\partial}{\partial r'}\left[ \frac{|r+r'| - |r-r'|}{r'r}\right]\nonumber\\
&=
\begin{cases}
{-\displaystyle k\,\frac{\partial r^{-1}}{\partial r'} = 0} & \mbox{for $r' < r$};\\
\vspace{-0.6cm} \\
{-\displaystyle k\,\frac{\partial r'^{-1}}{\partial r'} =
\frac{k}{r'^2}}&\mbox{for $r'>r$},
\end{cases}
\end{align}
which implies Eq.~(\ref{eq:4}).

This result can also be obtained by using the well known property
that the average of an electrostatic potential of a point charge
$Q$ over the surface of a sphere of radius $r$ equals the
potential at the center if the charge is outside the sphere, and
$Q/(4\pi\epsilon_0 r)$ if the charge is inside.\cite{griffiths2}
Stated mathematically, if $V(\bm r'') = k/\vert \bm r''\vert$,
then the average of $V$ over a spherical shell of radius $r$
around $\bm r'$ is
\begin{equation}
\overline{V_r}(\bm r') = k
\begin{cases}
r'^{-1} &\mbox{for $\vert \bm r'\vert >  r$};\\
r^{-1} &\mbox{for $\vert\bm r'\vert < r$}.
\end{cases}
\label{eq:a3}
\end{equation}
Averaging over spherical shells on both sides of the relation $\bm
G(\bm r') = -\nabla'V(\bm r')$ yields $\overline{\bm G_r}(\bm r')
= -\nabla'\,\overline{V_r}(\bm r')$. Using Eq.~(\ref{eq:a3}) in
this gives Eq.~(\ref{eq:4}).

\newpage
\begin{center}
\Large\underline{\bf Figures}
\end{center}
\begin{figure}[h]
\begin{center}
\includegraphics{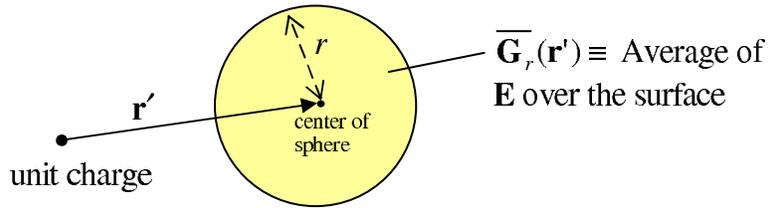}
\caption{\label{fig:averageG} Definition of $\overline{\bm
G}_r(\bm r')$: the average of the electric field of a unit charge
about a spherical surface of radius $r$, centered at $\bm r'$
relative to the charge.}
\end{center}
\end{figure}

\begin{figure}[h]
\includegraphics{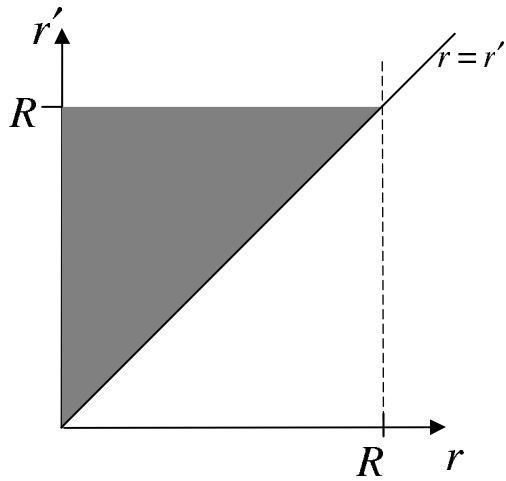}
\caption{\label{fig:2} Integration region for the limits $\int_0^R
dr \int_r^{R} dr'$ (grey region).  Integration over this region
can also be written as $\int_0^R dr' \int_0^{r'} dr.$}
\end{figure}

\begin{figure}[h]
\includegraphics{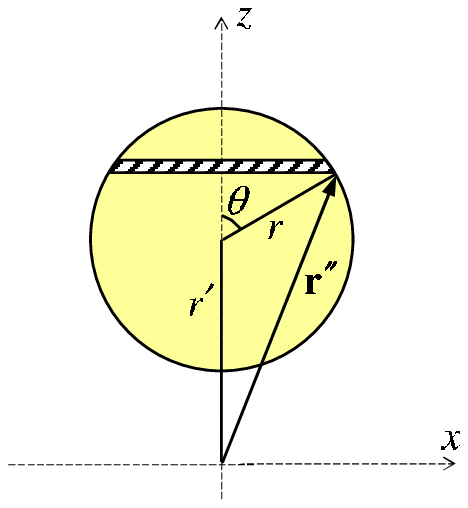}
\caption{\label{fig:3} Spherical surface over which $G_z$ is
averaged to obtain $\overline{ G_r}_{,z}$. The hatched strip
indicates the area element $dA''$ on the sphere with constant
$G_z(\theta) = k r_z''/\vert\bm r''\vert^3$ where $r''_z = r' +
r\cos\theta$ and $\vert \bm r''\vert = \sqrt{r'^2 + r^2 + 2 r r'
\cos\theta}$.}
\end{figure}

\begin{thebibliography}{99}

\bibitem{griffiths}
David J. Griffiths, {\em Introduction to Electrodynamics}
(Prentice-Hall, New Jersey, 1999) 3rd ed., pp. 173 -- 175.

\bibitem{jackson}
John D. Jackson, {\em Classical Electrodynamics} (Wiley, New York,
1999) 3rd ed., pp. 148 -- 150 and pp. 187 -- 188.

\bibitem{electric_only}
See {\em e.g.}, Wolfgang Pauli, {\em Electrodynamics} (MIT Press,
Cambridge, MA, 1973) pp. 37-39; Paul Lorrain, Dale P. Corson, and
Fran\c{c}ois Lorrain, {\em Electromagnetic Fields and Waves} (W.
H. Freeman, New York, 1988) 3rd ed., pp. 56-57; B. K. P. Scaife,
{\em Principles of Dielectrics} (Oxford University Press, Oxford,
1989), appendix F; Evaristo Riande and Ricardo Díaz-Calleja, {\em
Electrical Properties of Polymers} (Marcel Dekker, New York, 2004)
p.~44; Saunak Palit, {\em Principles of Electricity and Magnetism}
(Alpha Science International, Harrow, U.K., 2005) pp.~61-63; Tai
L. Chow, {\em Introduction to Electromagnetic Theory: A Modern
Perspective} (Jones and Bartlett, Boston, 2006) p.~86, problem 9.

\bibitem{both} See, {\em e.g.}, Ref. \onlinecite{griffiths}, pp.
156 -- 157, p. 253 and David J. Griffiths, {\em Instructor's
Solutions Manual: Introduction to Electrodynamics} (Prentice–Hall,
Englewood Cliffs, NJ, 1999), pp. 108 -- 109;  A. Z. Capri and P.
V. Panat, {\em Introduction to Electrodynamics} (Narosa, New
Delhi, 2002), pp.~155-158 and pp.~242-246; Ben Yu-Kuang Hu,
``Averages of static electric and magnetic fields over a spherical
region: A derivation based on the mean-value theorem," Am. J.
Phys. {\bf 68}, 1058--1060 (2000).

\bibitem{griffiths1} Ref. \onlinecite{griffiths}, p. 253, problem
5.57.

\bibitem{heavyside}  The Heavyside function is defined to be $\Theta(x) = 1$ for
$x > 0$, $\Theta(x) = 0$ for $x < 0$ and $\Theta(0) = \frac12$.

\bibitem{difference}  Note the difference between the average of a
vector field $\bm G$ over a  surface $\mathcal A$, $\int_{\mathcal
A} dA\ \bm G/\vert \mathcal A\vert$ (where $\vert\mathcal A\vert$
is the area of surface $\mathcal A$), and the flux of $\bm G$
through the surface, $\int_{\mathcal A} dA\ \hat{\bm n}\cdot\bm G$
(where $\hat{\bm n}$ is a unit vector perpendicular to area
element $dA$).  This paper deals with the former, while Gauss' law
pertains to the latter.

\bibitem{jackson1} See {\em e.g.}, Ref.~\onlinecite{griffiths}, p. 254; Ref.~\onlinecite{jackson}, p.
186.

\bibitem{griffiths2} See {\em e.g.}, Ref.~\onlinecite{griffiths},
pp. 114 -- 115.
\end{thebibliography}
\end{document}